\documentclass[10pt,letterpaper]{article}

\usepackage{graphicx}
\usepackage[utf8]{inputenc}
\usepackage[english]{babel}
\usepackage{relsize}
\usepackage[smaller,nolist]{acronym}
\usepackage{caption}
\usepackage{subcaption}
\usepackage{cleveref}
\usepackage{tabularx}
\usepackage{algorithm2e}
\usepackage{multirow}
\usepackage{listings}
\usepackage{xcolor}
\usepackage{url}
\usepackage{natbib}
\usepackage{amsfonts}

\usepackage[a4paper,top=3cm,bottom=2cm,left=2cm,right=2cm,marginparwidth=1.75cm]{geometry}

\definecolor{codegreen}{rgb}{0,0.6,0}
\definecolor{codegray}{rgb}{0.5,0.5,0.5}
\definecolor{codepurple}{rgb}{0.58,0,0.82}
\definecolor{backcolour}{rgb}{0.95,0.95,0.92}

\lstdefinestyle{mystyle}{
    backgroundcolor=\color{backcolour},   
    commentstyle=\color{codegreen},
    keywordstyle=\color{magenta},
    numberstyle=\tiny\color{codegray},
    stringstyle=\color{codepurple},
    basicstyle=\ttfamily\footnotesize,
    breakatwhitespace=false,         
    breaklines=true,                 
    captionpos=b,                    
    keepspaces=true,                 
    numbers=left,                    
    numbersep=5pt,                  
    showspaces=false,                
    showstringspaces=false,
    showtabs=false,                  
    tabsize=2
}
\begin{acronym}
    \acro{RS}{Recommender System}
    \acro{CF}{Collaborative Filtering}
    \acro{DL}{Deep Learning}
    \acro{MF}{Matrix Factorization}
    \acro{QoS}{Quality of Service}
    \acro{IoT}{Internet of Things}
    \acro{KNN}{K-Nearest Neighbors}
    \acro{PMF}{Probabilistic Matrix Factorization}
    \acro{NMF}{non-Negative Matrix Factorization}
    \acro{BNMF}{Bayesian non-Negative Matrix Factorization}
    \acro{DeepMF}{Deep Matrix Factorization}
    \acro{NCF}{Neural Collaborative Filtering}
    \acro{ML}{Machine Learning}
    \acro{MLP}{Multi-Layer Perceptron}
    \acro{VAE}{Variational AutoEncoder}
    \acro{VDeepMF}{Variational Deep Matrix Factorization}
    \acro{VNCF}{Variational Neural Collaborative Filtering}
    \acro{GAN}{Generative Adversarial Network}
    \acro{NLP}{Natural Language Processing}
    \acro{nDCG}{normalized Discounted Cumulative Gain}
    \acro{DCG}{Discounted Cumulative Gain}
    \acro{IDCG}{Ideal Cumulative Gain}
    \acro{MAE}{Mean Absolute Error}
    \acro{MSE}{Mean Squared Error}
\end{acronym}

\title{Deep Variational Models for Collaborative Filtering-based Recommender Systems}

\usepackage{authblk}
\author[1,2]{Jesús Bobadilla}
\author[1,2]{Fernando Ortega}
\author[1,2]{Abraham Gutiérrez}
\author[3,4,*]{Ángel González-Prieto}

\affil[1]{Departamento de Sistemas Informáticos, ETSI Sistemas Informáticos,\newline{} Universidad Politécnica de Madrid, Madrid, Spain.}
\affil[2]{KNODIS Research Group, ETSI Sistemas Informáticos,\newline{}Universidad Politécnica de Madrid, Madrid, Spain.}
\affil[3]{Departamento de Álgebra, Geometría y Topología, Facultad de Ciencias Matem\'aticas, \newline{}Universidad Complutense de Madrid, Madrid, Spain}
\affil[4]{Instituto de Ciencias Matem\'aticas (CSIC-UAM-UCM-UC3M), Madrid, Spain.}

\affil[*]{Corresponding author: angelgonzalezprieto@ucm.es}

\date{}

\begin{document}

\maketitle

\begin{abstract}
Deep learning provides accurate collaborative filtering models to improve recommender system results. Deep matrix factorization and their related collaborative neural networks are the state-of-art in the field; nevertheless, both models lack the necessary stochasticity to create the robust, continuous, and structured latent spaces that variational autoencoders exhibit.  On the other hand, data augmentation through variational autoencoder does not provide accurate results in the collaborative filtering field due to the high sparsity of recommender systems. Our proposed models apply the variational concept to inject stochasticity in the latent space of the deep architecture, introducing the variational technique in the neural collaborative filtering field. This method does not depend on the particular model used to generate the latent representation. In this way, this approach can be applied as a plugin to any current and future specific models.
The proposed models have been tested using four representative open datasets, three different quality measures, and state-of-art baselines. The results show the superiority of the proposed approach in scenarios where the variational enrichment exceeds the injected noise effect. Additionally, a framework is provided to enable the reproducibility of the conducted experiments.
\newline{}{\bf Keywords}: Recommender Systems, Collaborative Filtering, Variational Enrichment, Deep Learning.
\end{abstract}

\section{Introduction}
\label{sec:intro}

\Acp{RS} is an artificial intelligence field that provides methods and models to predict and recommend items to users (e.g.: films to persons, e-commerce products to costumers, services to companies, \ac{QoS}  to \ac{IoT} devices, etc.)~\citep{beel2013research}. Current popular \acp{RS} are Spotify, Netflix, TripAdvisor, Amazon, etc. \Acp{RS} are usually categorized attending to their filtering strategy, mainly demographic~\citep{bobadilla2021deep}, content-based~\citep{deldjoo2020recommender}, context-aware~\citep{kulkarni2020context}, social~\citep{shokeen2020study}, \ac{CF}~\citep{bobadilla2020deep,beel2013research} and filtering ensembles~\citep{forouzandeh2021presentation,ccano2017hybrid}. \Ac{CF} is the most accurate and widely used filtering approach to implement \acp{RS}. \Ac{CF} models have evolved from the \ac{KNN} algorithm to the \ac{PMF}~\citep{mnih2007probabilistic}, the \ac{NMF}~\citep{fevotte2011algorithms} and the \ac{BNMF}~\citep{hernando2016non}. Currently, deep learning research approaches are growing in strength: they provide improvement in accuracy compared to the \ac{ML}-based \ac{MF} models~\citep{rendle2020neural}. Additionally, deep learning architectures are usually more flexible than the \ac{MF}-based ones, introducing combined deep and shallow learning~\citep{he2017neural}, integrated content-based ensembles~\citep{narang2018deep}, generative approaches~\citep{bobadilla2020deepfair,gao2021recommender}, among others.

\Ac{DeepMF}~\citep{xue2017deep} is a neural network model that implements the popular \ac{MF} concept. \Ac{DeepMF} was designed to take as input a user-item matrix with explicit ratings and nonpreference implicit feedback, although current implementations use two embedding layers whose inputs are, respectively, user and items. The experimental results evidence the \ac{DeepMF} superiority over the traditional approaches based on \ac{ML}-focused \ac{RS}, particularly the most used \ac{MF} models: \ac{PMF}, \ac{NMF}, and \ac{BNMF}. Currently, \ac{DeepMF} is a popular model that is rapidly replacing the traditional \ac{MF} models based on classical \ac{ML}. Additionally, \Ac{DeepMF} has been used in the \ac{RS} field to combine social behaviors (clicks, ratings,…) with images~\citep{wen2018visual}, and a social trust-aware \ac{RS} has been implemented by using \ac{DeepMF} to extract features from the user-item rating matrix for improving the initialization accuracy~\citep{wan2020deep}. \ac{QoS} predictions have also been addressed by using \ac{DeepMF}~\citep{zou2020ndmf}. To learn attribute representations, a \ac{DeepMF} model has been used that creates a low-dimensional representation of a dataset that lends itself to a clustering interpretation~\citep{trigeorgis2016deep}. Finally, the classical matrix completion task has been addressed by using the \ac{DeepMF} approach~\citep{fan2018matrix}.

\begin{figure}[h]
    \centering
    \includegraphics[width=0.95\textwidth]{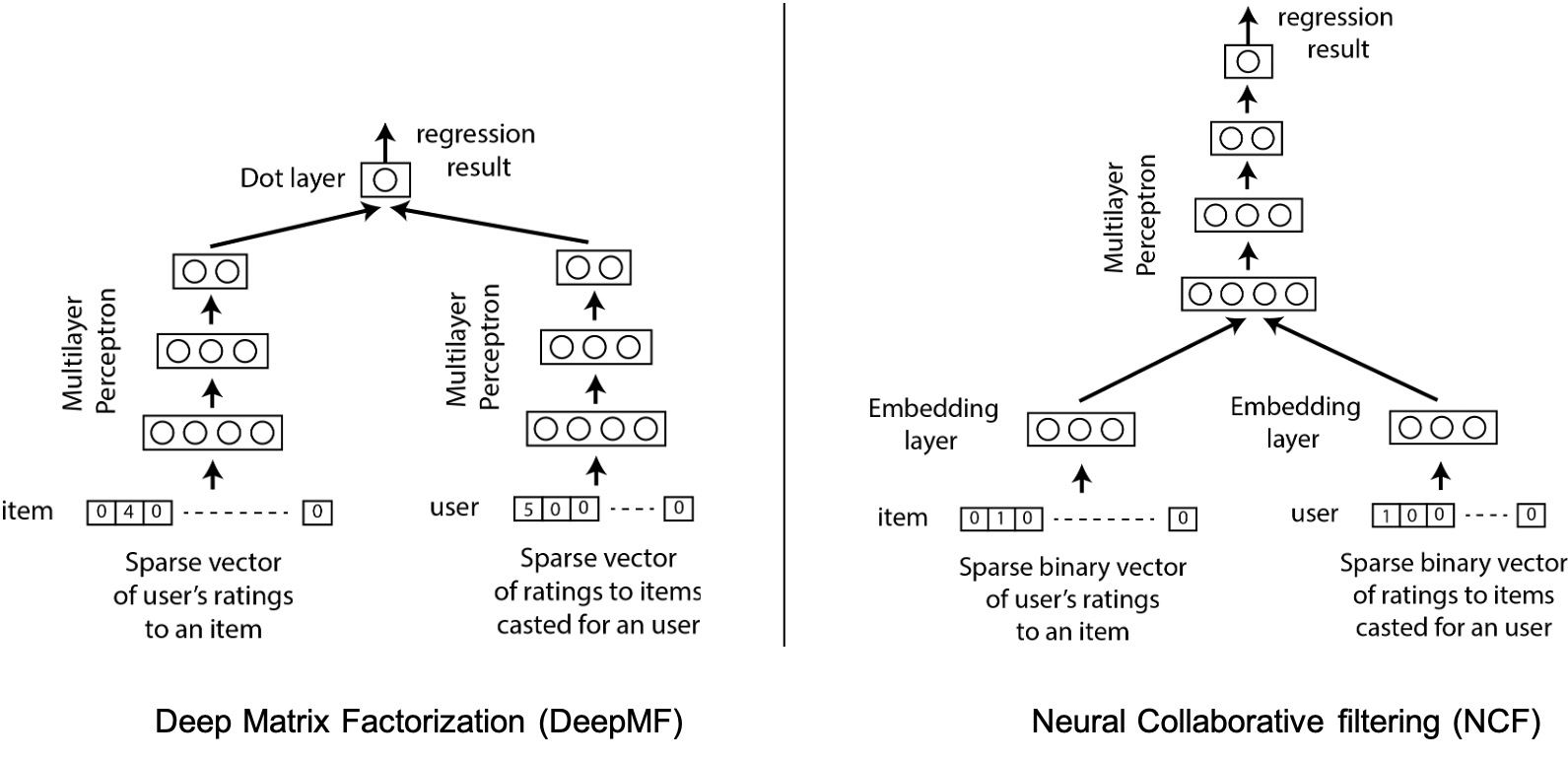}
    \caption{Deep Matrix Factorization (DeepMF) versus Neural Collaborative Filtering (NCF).}
    \label{fig:deepmf-vs-ncf}
\end{figure}

The not so widely spread \ac{NCF} model~\citep{he2017neural} may be seen as an augmented \ac{DeepMF} model, where deeper layers are added to the ‘\texttt{Dot}’ one. Additionally, the ‘\texttt{Dot}’ layer can be replaced by a ‘\texttt{Concatenate}’ layer. \Cref{fig:deepmf-vs-ncf} shows the explained concepts. \Ac{NCF} slightly outperforms the \ac{DeepMF} accuracy results, but it increases the required runtime to train the model and to run the forward process: it is necessary to execute the `extra' \ac{MLP} on top of the `\texttt{Dot}' or `\texttt{Concatenate}' layers. Moreover, compared to \ac{DeepMF}, the \ac{NCF} architecture adds new hyper-parameters to set: mainly the number of hidden layers (depth) and their size (number of neurons in each layer) of the \ac{MLP} architecture.

\begin{figure}[h]
    \centering
    \includegraphics[width=0.7\textwidth]{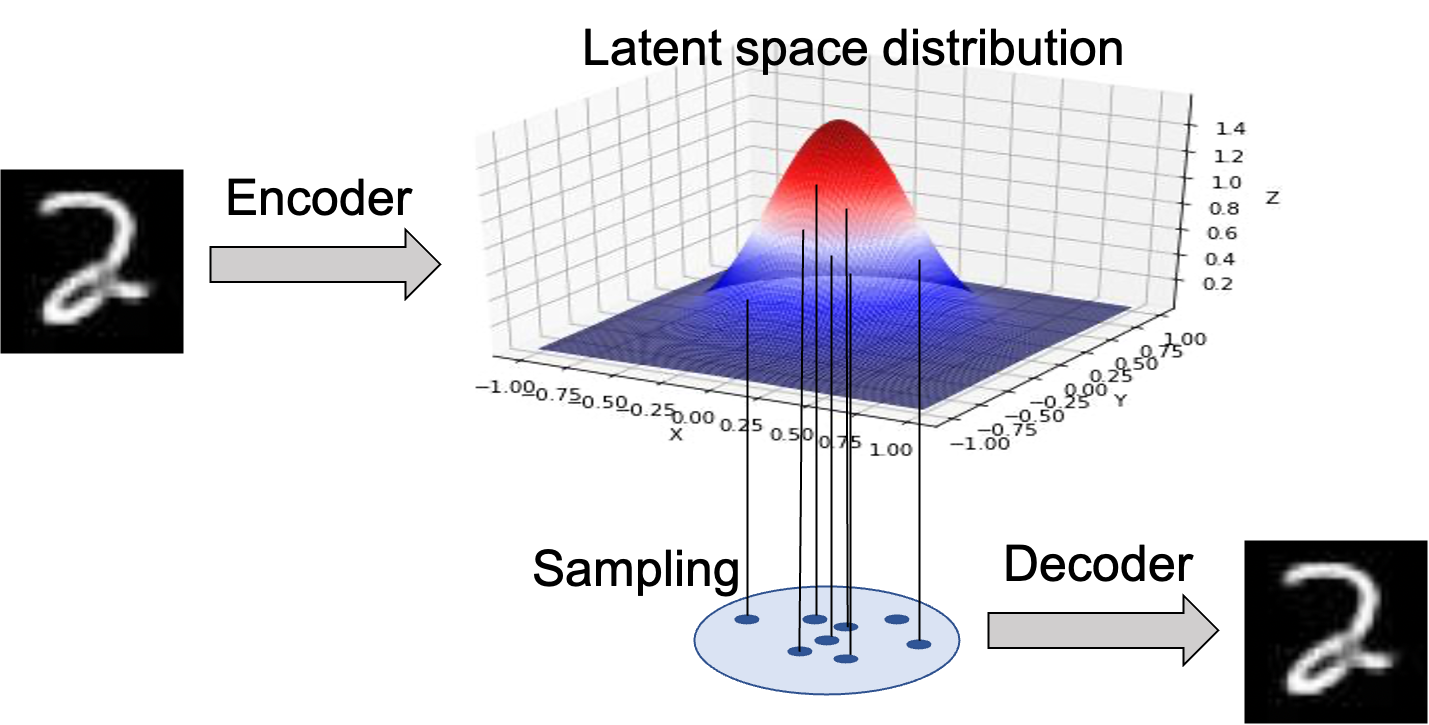}
    \caption{Variational autoencoder model.}
    \label{fig:vae}
\end{figure}

\begin{figure}[h]
    \centering
    \includegraphics[width=0.7\textwidth]{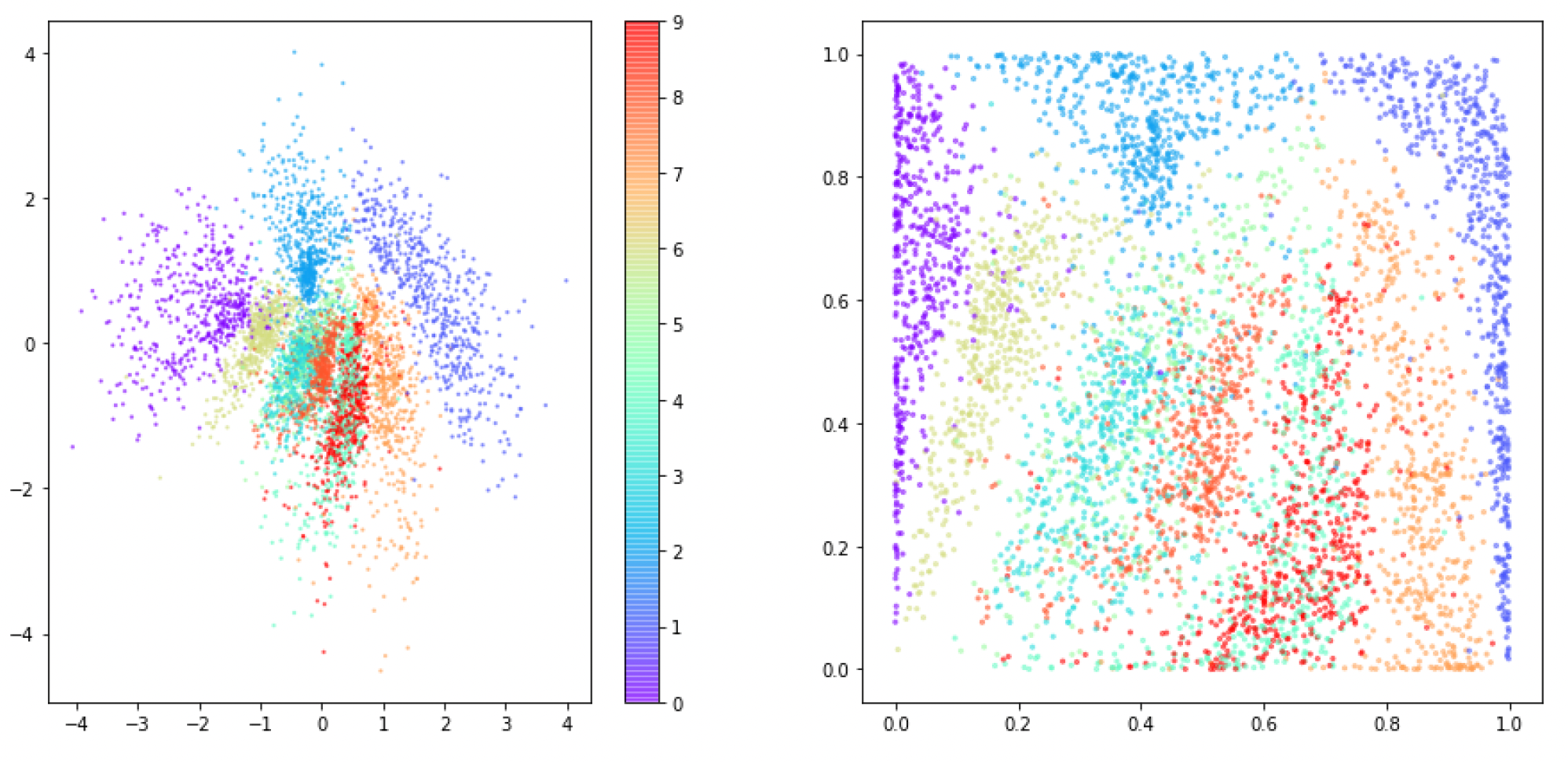}
    \caption{Representation of an autoencoder latent space for the MNIST dataset (left side) versus the equivalent \ac{VAE} latent space representation (right side).}
    \label{fig:vae-mnist}
\end{figure}

In a different setting, \acp{VAE} act as regular autoencoders. They aim to compress the input raw values into a latent space representation by means of an encoder neural network, whereas the decoder neural network makes the opposite operation seeking to decompress from latent space to output raw values. The main difference between classical autoencoders and \acp{VAE} is the latent space design, meaning, and operation. Classical autoencoders do not generate structured latent spaces, whereas \acp{VAE} introduce a statistical process that forces them to learn continuous and structured latent spaces. In this way, \acp{VAE} turn the samples into parameters of a statistical distribution, usually the means and variance of a Gaussian distribution. This concept is illustrated in \Cref{fig:vae}. From the parameters in the multivariate distribution, we draw a random sample and a latent space sample is obtained for each training input (center of \cref{fig:vae}). The stochasticity of the random sampling improves the robustness and forces the encoding of continuous and meaningful latent space representations, as it can be seen in \cref{fig:vae-mnist}, where it is shown the difference between a regular autoencoder latent space representation and its equivalent \ac{VAE} one.   

Due to their properties, \acp{VAE} have been used as generative deep learning models in the image processing field. Reconstruction of a multispectral image has been performed by means of a \ac{VAE}~\citep{liu2020multispectral} that parameterizes the latent space of Gaussian distribution parameters. \acp{VAE} have been also used to create superresolution images as in \citet{liu2020unsupervised}, where a model is proposed to encode low-resolution images in a dense latent space vector that can be decoded for target high resolution image denoising. The blur image problem using \ac{VAE} is tackled in \citet{liu2020photo} by adding a conditional sampling mechanism that narrows down the latent space, making it possible to reconstruct high resolution images. Moreover, in \citet{zhang2021online}, the authors propose a flexible autoencoder model able to adapt to varying data patterns with time. By importing the \ac{VAE} concept from image processing, several papers have used these models to improve \ac{RS} results. For instance, denoising and variational autoencoders are tested in \citet{liang2018variational}, where the authors reported the superiority of the \ac{VAE} option against other models, or in \citet{nisha2019social}, where variational autoencoders are combined with social information to improve the quality of the recommendations.

The aim of this paper is to propose a neural architecture that joins the best of the \ac{DeepMF} and \ac{NCF} models with the \ac{VAE} concept. This novel models will be called, respectively, \ac{VDeepMF} and \ac{VNCF}. In contrast with the autoencoder and \ac{GAN} approaches in the \ac{CF} field~\citep{gao2021recommender}, we shall not use the generative decoder stage and we maintain the regression output layer presented in the \ac{DeepMF} and the \ac{NCF} models. The main advantage in the use of the \ac{VAE} operation is the robustness that it confers to the latent representation. This robustness can be seen by observing \cref{fig:vae-mnist}. If we consider each dot drawn as a train sample representation in the latent space, then test samples are most likely to be correctly classified in the  \ac{VAE} model (right graph in \cref{fig:vae-mnist}) than being correctly classified in the regular autoencoder model (left graph in \cref{fig:vae-mnist}). In short, the variational approach stochastically `spreads' the samples in the latent space, improving the chances of classifying correctly the training samples. 

\begin{figure}[h]
    \centering
    \includegraphics[width=0.7\textwidth]{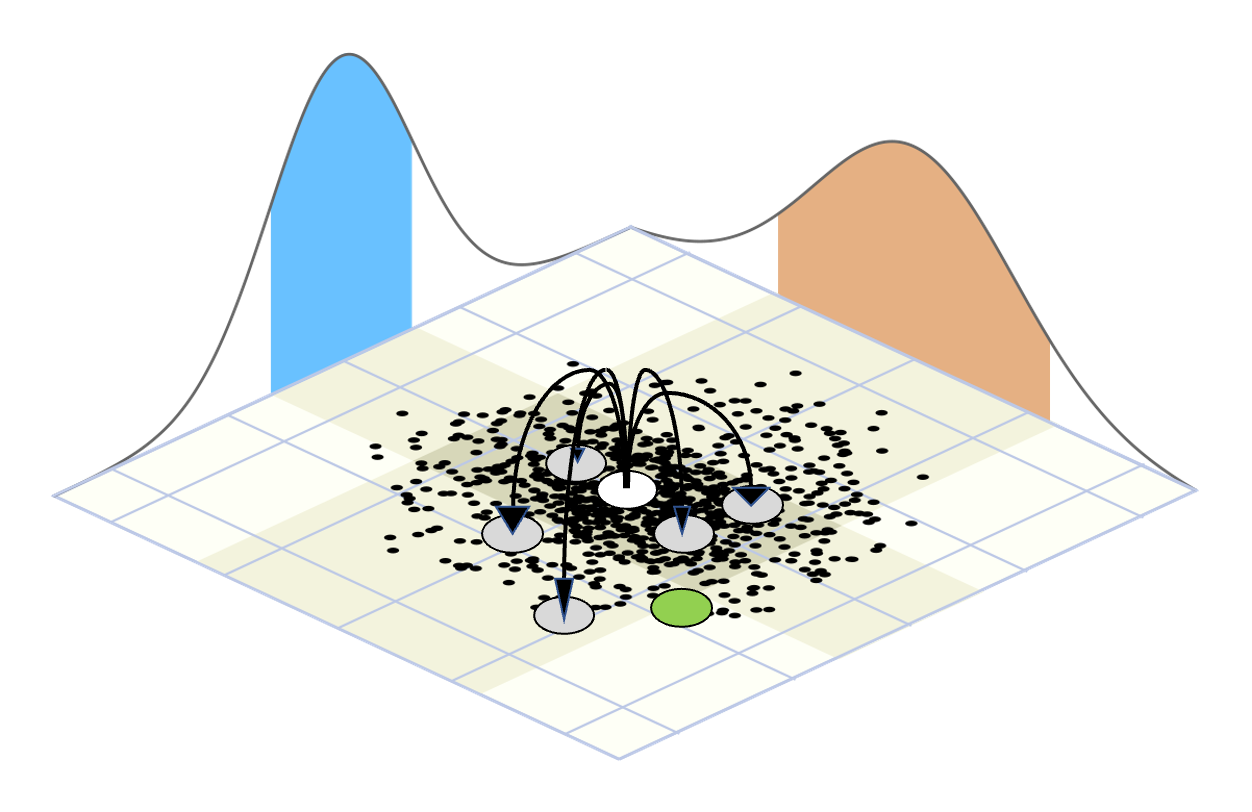}
    \caption{Latent space representation of the proposed variational model. From the learnt means and variances of the multivariate Gaussian distribution, a random sampling process is run to spread the latent space sample values (gray circles) that will help to accurately predict the unknown sample rating values (green circle).}
    \label{fig:latent-space}
\end{figure}

In our proposed \ac{RS} \ac{CF} scenario, we expect that rating values can be better predicted when a variational latent space has been learnt, because this space covers a wider, more robust, and more representative latent area. Whereas with a traditional autoencoders each sample would be coded as a value in the latent space (white circle in \cref{fig:latent-space}), the \ac{VAE} encodes the parameters of a multivariate distribution (e.g.\ mean and variance of both the blue and the orange Gaussian distributions in \cref{fig:latent-space}). From the learnt distribution parameters, random sampling is carried out to generate stochastic latent space values (gray circles in \cref{fig:latent-space}). Each epoch in the learning process generates a new set of latent space values. Once the proposed model has been trained, when a $\langle \textrm{user, item}\rangle$ tuple is presented to the model, the obtained latent space value (green circle in \cref{fig:latent-space}) can be better predicted in the \ac{VAE} scenario than in the regular autoencoder scenario: the random sampled values (gray circles) of the enriched latent space will help to associate the predicted sample (green circle) with their associated training samples (white circle), making the prediction process much more robust and accurate.

Current \ac{CF}-based variational autoencoders usually obtain raw augmented data: mainly synthetic ratings from user to items or generated relevant versus not relevant votes from users to items~\citep{liang2018variational,gao2021recommender}. This strategy forces us to sequentially run two separated models: the generative model (\ac{GAN} or \ac{VAE}) that provides augmented data, and the regression \ac{CF} model that makes predictions and recommendations. This approach presents three main drawbacks: 1) complexity, as two separate models are necessary, 2) large time consumption, and 3) sparsity management. As we will explain deeper in the following section, our proposed model does not generate raw augmented data. On the contrary, its innovation is based on the use of a single model to internally manage both augmentation and prediction aims. Particularly significant is the way in which the proposed model addresses the sparsity problem: we do not make augmentation on the sparse raw data (ratings cast from users to item), but an internal ‘augmentation’ process in the dense latent space of the model (\cref{fig:vae-mnist,fig:latent-space}). Each sample that is randomly generated from the latent space feeds the model regression layers. Thereby, we propose a model that first generates stochastic variational samples in a dense latent space, and then these generated samples act as inputs of the regression stage of the model. 

To test these ideas, the hypothesis considered in this paper is that the augmented samples will be more accurate and effective if they are generated in an inner and dense latent space rather than in a very sparse input space. It is important to realize that enriching the inner latent space can improve the recommendation results, but it also injects noise to the latent space that may potentially worsen the results. It is expected that the proposed approach will work better with poor latent spaces, whereas when it is applied to rich spaces, the spurious entropy added by the variational stage could worsen recommendations. Thus, medium-size \ac{CF} datasets, or large and complex ones are better candidates to improve their results when the variational proposal is applied, whereas large datasets with predictable data distributions will probably not benefit from the noise injection of the variational architecture.
    
The rest of the paper has been structured as follows. In \cref{sec:model}, the proposed model is explained. \Cref{sec:experiments} shows the experiments’ design, results and their discussions. Finally, \cref{sec:conclusions} contains the main conclusions of the paper and the future works.

\section{Proposed model}
\label{sec:model}

The proposed neural architecture will mix the \ac{VAE} and the \ac{DeepMF} (or the \ac{NCF}) models. From the \ac{VAE} we take the encoder stage and its variational process, and from the \ac{DeepMF} or the \ac{NCF} model we use its regression layers. This is an innovative approach in the \ac{RS} field, since the \ac{VAE} and \ac{GAN} neural networks have only been used as a posteriori stage to make data augmentation, i.e.\ to obtain enriched input datasets to feed the \ac{CF} \ac{DeepMF} or \ac{NCF} models. Hence, the traditional approach needs to separately train two models, first the \ac{VAE} and then the \ac{DeepMF}/\ac{NCF} networks.

\begin{figure}[h]
    \centering
    \includegraphics[width=0.95\textwidth]{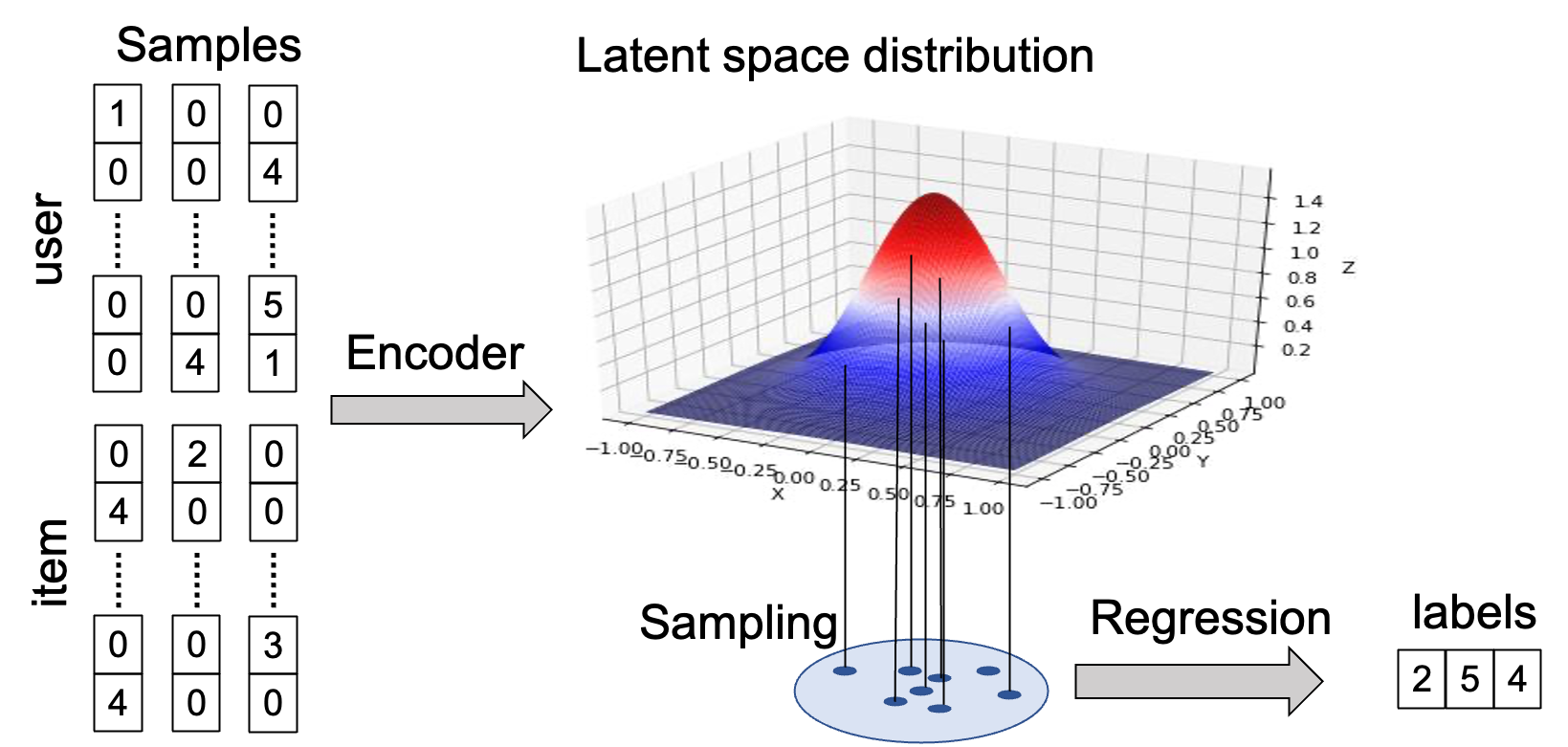}
    \caption{Proposed VDeepMF/NCF approach. CF samples are encoded in the latent space by means of a variational process and then predictions are obtained by using a regression neural network.}
    \label{fig:roposed-approach}
\end{figure}

In sharp contrast, our proposed approach efficiently joins the \ac{VAE} and the Deep \ac{CF} regression concepts to obtain improved predictions with a single training process. In the learning stage, the training samples feed the model (left hand side of \cref{fig:roposed-approach}). Each training sample consists of the tuple $\langle \textrm{user, item, rating}\rangle$ (rating casted by the user to the item). In the \ac{DeepMF}/\ac{NCF} architecture, each user is represented by his/her vector of voted ratings, and each item is represented by its vector of received ratings. The model learns the ratings (third element in the tuples) casted by the users to the items (first and second elements in the tuples). In other words, the ratings are outputs of the neural network (right hand side of \cref{fig:roposed-approach}).

\subsection{Formalization of the model}
\label{subsec:method-formalization}

\begin{figure}[h]
    \centering
    \includegraphics[width=0.95\textwidth]{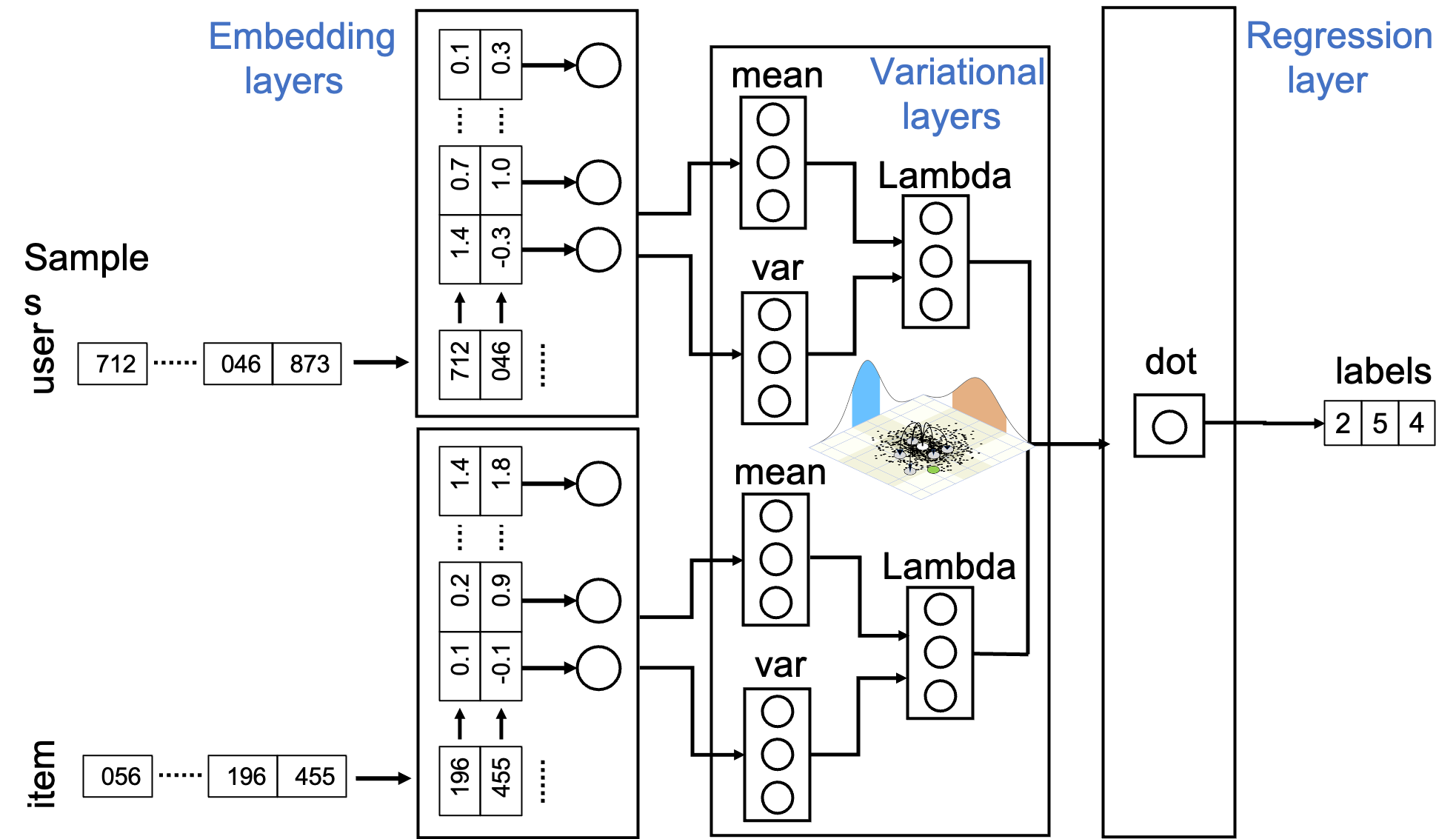}
    \caption{Proposed VDeepMF architecture. The NCF architecture will have identical ‘Embedding’ and ‘Variational’ layers to the VDeepMF one; it will just replace the ‘Dot’ layer for a ‘Concatenate’ layer, followed by an MLP.}
    \label{fig:roposed-architecture}
\end{figure}

The architectural details of the proposed models are shown in \cref{fig:roposed-architecture}. For simplicity, only the \acf{VDeepMF} architecture is shown in this figure. The corresponding model for \ac{NCF}, named \acf{VNCF}, is analogous to the \ac{VDeepMF} one: it has the same `\texttt{Embedding}' and `\texttt{Variational}' layers and we should only replace the `\texttt{Dot}' layer of \ac{DeepMF} by a `\texttt{Concatenate}' layer followed by a \ac{MLP}.

To fix the notation, let us suppose that our dataset contains $U$ users and $I$ items. In general, the aim of any deep learning model for \ac{CF}-based prediction is to train a (stochastic) neural network that implements a function
$$
    h: \mathbb{R}^U \times \mathbb{R}^I \to \mathbb{R}. 
$$
This function $h$ operates as follows. Let us codify the $u$-th user of the dataset (resp.\ the $i$-th item) using one-hot-encoding as the $u$-th canonical basis vector $\textbf{e}_u$ (resp.\ the $i$-th canonical basis vector $\textbf{e}_i$). Then, $h(\textbf{e}_u, \textbf{e}_i) \in \mathbb{R}$ seeks to predict the score that the $u$-th user would assign to the $i$-th item. To train this function $h$, in the learning phase the neural network is fed with a set of training tuples $\langle \textrm{u, i, r} \rangle$ of user $u$ that rated item $i$ with a score $r$ and the function $h$ is trained to fit $h(\textbf{e}_u, \textbf{e}_i) = r$. 

Our proposal for the \ac{VDeepMF} consist on decomposing $h$ has a combination of a `\texttt{Embedding}', followed by a `\texttt{Variational}' stage and a final `\texttt{Dot}' layer, as shown in \cref{fig:roposed-architecture}). The first `\texttt{Embedding}' layer (left hand side of \cref{fig:roposed-architecture}) is borrowed from the natural language processing field~\citep{he2017neural}. The idea is that this layers provides a fast translation of users and items into their respective representations in the latent spaces. To be precise, this layer implements a function $\texttt{Embedding}$ that maps a pair $(\textbf{e}_u, \textbf{e}_i)$ into a pair of dense vectors $(v_u, w_i) \in \mathbb{R}^{L} \times \mathbb{R}^L$ that represents the $u$-th user and the $i$-th item, being $L > 0$ the dimension of the representations.

It is worth mentioning that, even though from a conceptual point of view the `\texttt{Embedding}' layer is a regular \ac{MLP} dense layer, to save time and space, these `\texttt{Embedding}' layers are typically implemented through lookup tables. In this way, instead of feeding the network with the one-hot encoding of the user $u$ (resp. the item $i$), we input it via its ID as user (resp.\ as item). The lookup table efficiently recovers the $u$-th (resp.\ $i$-th) column of the embedding matrix that contains $v_u$ (resp.\ $w_i$) so that the translation can be conducted in a more efficient way than with a standard \ac{MLP} layer by exploiting the sparsity of the input.

The variational process is carried out by the `\texttt{Variational}' stage (labeled as `variational layers' at the middle of \cref{fig:roposed-architecture}). From the latent space representation $(v_u, w_i) \in \mathbb{R}^{L} \times \mathbb{R}^L$ of the the $u$-th user and the $i$-th item, two separated dense layers return the mean and variance parameters of two gaussian multivariate distribution. In this way, if fix a latent space dimension $K>0$, the first part of this `\texttt{Variational}' stage (left part of the middle rectangle of \cref{fig:roposed-architecture}) computes a map
$$
    (v_u, w_i) \mapsto (\mu_1(v_u), \sigma_1^2(v_u), \mu_2(w_i), \sigma^2(w_i)) \in \mathbb{R}^{4K},
$$
where $\mu_1(v_u), \mu_2(w_i)$ represent the means of the associated gaussian distributions to the user and the item respectively, and $\sigma_1^2(v_u), \sigma^2(w_i)$ their variance. Thus, the output of the `\texttt{Variational}' stage (left right of the middle rectangle of \cref{fig:roposed-architecture}) is a pair of random vectors $(P_{\mu_1(v_u),\sigma_1^2(v_u)}, Q_{\mu_2(w_i),\sigma^2(w_i)})$ where
$$
    P \sim \mathcal{N}(\mu_1(v_u), \textrm{diag}\,\sigma_1^2(v_u)), \qquad Q \sim \mathcal{N}(\mu_2(w_u), \textrm{diag}\,\sigma_2^2(w_i)).
$$
Here, $\mathcal{N}(\mu, \Sigma)$ denotes a $K$-dimensional multivariate normal distribution of mean vector $\mu$ and diagonal covariance matrix $\Sigma$, so that our covariance matrix is always diagonal. Each time a sample is drawn, the `\texttt{Variational}' stage thus returns a pair $(p, q) \in \mathbb{R}^K \times \mathbb{R}^K$, which represents the stochastic latent representations associated to $(v_u, w_i)$.

The final `\texttt{Dot}' layer (labeled as `regression layer' at right hand side of \cref{fig:roposed-architecture}) in the \ac{VDeepMF} model is very simple. It is a linear layer that simply computes the dot product of the latent vectors $p$ and $q$. Therefore
$$
\texttt{Dot}(p,q) = p \cdot q.
$$
In the case of \ac{VNCF}, this simple layer is replaced by a fully connected \ac{MLP} that extracts non-linear relations from $p$ and $q$.

Therefore, summarizing the process, the proposed \ac{VDeepMF} model $h$ computes
$$
    h(\textbf{e}_u, \textbf{e}_i) = \texttt{Dot} \circ \texttt{Variational} \circ \texttt{Embedding}(\textbf{e}_u, \textbf{e}_i) =  P_{\mu_1(v_u),\sigma_1^2(v_u)} \cdot Q_{\mu_2(w_i),\sigma^2(w_i)}.
$$
This is a random variable that, when sampled, returns a natural number that should be interpreted as the predicted rating by $h$ for the user $u$ regarding item $i$.


\subsection{Implementation of the model}

The model described in Section \ref{subsec:method-formalization} has been implemented in Keras \citep{chollet2015keras}, a widely used Python library for deep learning and neural computing. 
For the sake of reproducibility, the code framework that implements the architecture shown in \cref{fig:roposed-architecture} (both in their \ac{VDeepMF} and \ac{NCF} versions) and the experiments explained in the next section is available at the GitHub repository\footnote{\url{https://github.com/KNODIS-Research-Group/deep-variational-models-for-collaborative-filtering}}.

Additionally, as an example, Listing \ref{lst:vdeepmf} shows the source code of the proposed \ac{VDeepMF} kernel: lines 8 to 13 implement the user side of the \cref{fig:roposed-architecture} architecture, whereas lines 15 to 20 do the same job on the item side of \cref{fig:roposed-architecture}. Please note the use of the Keras Embedding layers in lines 9 and 16. Lines 10-12 and 17-19 carry out the `\texttt{Variational}' stage. In particular, both the user and the item \texttt{Lambda} layers (lines 12 and 19) run the variational process. They use the \texttt{sampling} function (lines 3 to 6) to combine the mean and variance latent values, which make use of the Keras backend \texttt{random\_normal} procedure to implement the stochasticity (line 5). Finally, the latent values of users and items are combined by means of the `\texttt{Dot}' layer (line 22) to produce the final output.

\begin{lstlisting}[language=Python, caption=VDeepMF kernel code., numberblanklines=false, label=lst:vdeepmf]
latent_dim=5; batch_size = 32

def sampling(args):
    z_mean, z_var = args
    epsilon = K.random_normal(shape=(batch_size, latent_dim), mean=0., stddev=1)
    return z_mean + K.exp(z_var) * epsilon

user_input = Input(shape=[1])
user_embedding = Embedding(num_users, latent_dim)(user_input)
user_embedding_mean = Dense(latent_dim)(user_embedding)
user_embedding_var = Dense(latent_dim)(user_embedding)
user_embedding_z = Lambda(sampling)([user_embedding_mean, user_embedding_var])
user_vec = Flatten()(user_embedding_z)

item_input = Input(shape=[1])
item_embedding = Embedding(num_items, latent_dim)(item_input)
item_embedding_mean = Dense(latent_dim)(item_embedding)
item_embedding_var = Dense(latent_dim)(item_embedding)
item_embedding_z = Lambda(sampling)([item_embedding_mean, item_embedding_var], latent_dim)
item_vec = Flatten()(item_embedding_z)

dot = Dot(axes=1)([item_vec, user_vec])

VDeepMF = Model([user_input, item_input], dot)
\end{lstlisting}

\section{Empirical evaluation}
\label{sec:experiments}

In this section, we describe the empirical experiments carried out to evaluate the performance of the variational approach in the \ac{DeepMF} and \ac{NCF} models.

\subsection{Experimental setup}

The experimental evaluation has been performed over four different datasets to measure the performance of the proposed method over different environments. The selected datasets are: FilmTrust~\citep{guo2013novel}, an small dataset that contains the ratings of thousands of items to movies;  MovieLens 1M~\citep{harper2015movielens}, the gold standard dataset in \ac{CF}-based \ac{RS}; MyAnimeList~\citep{myanimelist}, a dataset extracted from Kaggle\footnote{\url{www.kaggle.com}} that contains the ratings of thousands of users to anime comics; and Netflix~\citep{bennett2007netflix}, a popular dataset with hundred of millions ratings used in the Neflix Prize competition. \Cref{tab:datasets} show the main parameters of these datasets. The corpus of these datasets has been randomly splitted into training ratings (80\% of the ratings) and test ratings (20\% of the ratings).

\begin{table}[h]
\centering
\begin{tabular}{|l|l|l|l|l|l|}
    \hline
    Dataset     & Number of users & Number of items & Number of ratings & Scores     & Sparsity \\ \hline
    FilmTrust   & 1,508           & 2,071           & 35,494            & 0.5 to 4.0 & 98.86\%  \\ \hline
    MovieLens   & 6,040           & 3,706           & 1,000,209         & 1 to 5     & 95.53\%  \\ \hline
    MyAnimeList & 69,600          & 9,927           & 6,337,234         & 1 to 10    & 99.08\%  \\ \hline
    Netflix     & 480,189         & 17,770          & 100.480.507       & 1 to 5     & 98.82\%  \\ \hline
\end{tabular}
\caption{Main parameters of the datasets used in the experiments.}
\label{tab:datasets}
\end{table}

The evaluation of the proposed method has been analyzed from three different points of view: the quality of the predictions, the quality of the recommendations, and the quality of the recommendation lists.

To measure the quality of the predictions, we have compared the real rating $r_{u,i}$ of an user $u$ to an item $i$ of the test split $R^{test}$ with the predicted one, $\hat{r}_{u,i}$. These comparison has been carried out in three ways: using the \ac{MAE} as in \cref{eq:mae}, using the \ac{MSE} as in \cref{eq:mse} and computing the proportion of the explained variance $R^2$ as in \cref{eq:r2}. Notice that, in (\cref{eq:r2}), $\bar{r}$ denotes the mean of the ratings contained in the test split.

\begin{equation} \label{eq:mae}
    \textrm{MAE} = \frac{1}{\#R^{test}} \sum_{\langle u,i \rangle \in R^{test}} | r_{u,i} - \hat{r}_{u,i} |,
\end{equation}

\begin{equation} \label{eq:mse}
    \textrm{MSE} = \frac{1}{\#R^{test}} \sum_{\langle u,i \rangle\in R^{test}} \left( r_{u,i} - \hat{r}_{u,i} \right)^2,
\end{equation}

\begin{equation} \label{eq:r2}
    R^2 = 1 - \frac{{\displaystyle \sum_{\langle u,i \rangle \in R^{test}} \left( r_{u,i} - \hat{r}_{u,i} \right)^2}}{{\displaystyle \sum_{\langle u,i \rangle \in R^{test}} \left( r_{u,i} - \bar{r} \right)^2}}.
\end{equation}

To measure the quality of the recommendations, we have analyzed the impact of the top $N$ recommended items to the user $u$, collected in the list $T_u^N$. Using precision (\cref{eq:precision}), we measure the proportion of relevant recommendations (i.e.\ the user rated the item with a rated equal or greater than a threshold $\theta$) among the top $N$. Here $U$ denotes the set of user in the test split. In a similar vein, using recall (\cref{eq:recall}), we measure the proportion of the test items rated by the user $u$, $R_u^{test}$, that were relevant to him or her and were included into the recommended items $T_u^N$. For the conducted experiments, the used thresholds are $\theta=3$ for FilmTrust, $\theta=4$ for MovieLens and Netflix, and $\theta=8$ for MyAnimeList. 

\begin{equation} \label{eq:precision}
    \textrm{Precision} = \frac{1}{\#U} \sum_{u \in U} \frac{\lbrace i \in T_u^N | r_{u,i} \geq \theta \rbrace}{N},
\end{equation}

\begin{equation} \label{eq:recall}
    \textrm{Recall} = \frac{1}{\#U} \sum_{u \in U} \frac{\lbrace i \in T_u^N | r_{u,i} \geq \theta \rbrace}{\lbrace i \in R_u^{test} | r_{u,i} \geq \theta \rbrace}.
\end{equation}

Finally, to measure the quality of the recommendation lists we use the \ac{nDCG}. Suppose that the recommendation list of the user $u$, $T_{u}^N$, is sorted decreasingly so that the items predicted as more relevant are placed in the first positions. Given $i \in T_{u}^N$, let $\textrm{pos}_{T_{u}^N}(i)$ be the position of the item $i$ in the recommendation list. Analogously, suppose that the real top $N$ recommendations to user $u$, $R_u^N$, as sorted decreasingly and denote by $\textrm{pos}_{R_{u}^N}(i)$ the position of the item $i \in R_u^N$ in the list. In this setting, the \ac{DCG} and the \ac{IDCG} of the user $u \in U$ are defined as in (\cref{eq:dcg-idcg}).
\begin{equation}\label{eq:dcg-idcg}
    \textrm{DCG}_u = \sum_{i \in T_{u}^N} \frac{2^{r_{u,i}}-1}{\log_2 \left(\textrm{pos}_{T_{u}^N}(i)+1\right)}, \qquad \textrm{IDCG}_u = \sum_{i \in R_{u}^N} \frac{2^{r_{u,i}}-1}{\log_2 \left(\textrm{pos}_{R_{u}^N}(i)+1\right)}.
\end{equation}
In this way, \ac{nDCG} is given by the mean of the ratio between \ac{DCG} and \ac{IDCG} as in (\cref{eq:ndcg}).


\begin{equation} \label{eq:ndcg}
    \textrm{nDCG} = \frac{1}{\#U} \sum_{u \in U} \frac{\textrm{DCG}_u}{\textrm{IDCG}_u}.
\end{equation}



Due to the stochastic nature of the variational embedded space of the proposed method, the test predictions used to evaluate the proposed method have been computed as the average of the $10$ predictions performed for each pair of user $u$ and item $i$.

\subsection{Experimental results}

\Cref{tab:prediction-error} includes the quality of the predictions performed by the proposed model. Best values for each dataset are highlighted in bold. \Cref{tab:mae} contains the \ac{MAE} (\cref{eq:mae}), \cref{tab:mse} contains the \ac{MSE} (\cref{eq:mse}), and \cref{tab:r2} contains the $R^2$ score (\cref{eq:r2}). We can observe that the proposed variational approach improves the prediction capability of \ac{DeepMF} in all datasets except of Netflix and reports worse predictions when it is applied to \ac{NCF}.

We justify these results by taking into account the features of the deep learning models used and the properties of each dataset. On the one hand, the larger the size of the dataset, the less necessary it is to enrich the votes with the proposed variational approach. In other words, when the dataset is small, the amount of Shannon entropy \citep{shannon1949mathematical} that it contains might be quite limited. By using a variational method to generate new samples, we add some extra entropy that enriches the dataset, giving the chance to the regressive part of exploiting this extra data. However, large datasets usually present a large entropy in such a way that the regressive models can effectively extract very subtle information from them. In this setting, if we add a variational stage, instead of adding new relevant variability to the dataset, we only add noise that muddies the underlying patterns. For this reason, the variational approach is of no benefit in huge datasets like Netflix.

On the other hand, the \ac{NCF} model is more complex than the \ac{DeepMF} one, so data enrichment has less impact for complex models that are able to find more sophisticated relationships between data than simpler models. In fact, based on these results, we can assert that including the variational approach into a simple model such as \ac{DeepMF} is equivalent to using a more complex model such as \ac{NCF}.

\begin{table}[h]
    \begin{subtable}[h]{\textwidth}
        \centering
        \begin{tabular}{|l|l|l|l|l|}
        \hline
                & FilmTrust       & MovieLens       & MyAnimeList     & Netflix         \\ \hline
        VDeepMF & \textbf{0.6344} & \textbf{0.6815} & \textbf{0.8805} & 0.7246          \\ \hline
        DeepMF  & 0.7601          & 0.7036          & 0.9050          & \textbf{0.6834} \\ \hline
        VNCF    & 0.6730          & 0.6934          & 0.9142          & 0.7441          \\ \hline
        NCF     & \textbf{0.6308} & 0.6940          & 0.8905          & 0.6905          \\ \hline
        \end{tabular}
        \caption{Mean Absolute Error. The lower the better.}
        \label{tab:mae}
    \end{subtable}
    
    \vspace{0.3cm}
    
    \begin{subtable}[h]{\textwidth}
        \centering
        \begin{tabular}{|l|l|l|l|l|}
        \hline
                & FilmTrust       & MovieLens       & MyAnimeList     & Netflix         \\ \hline
        VDeepMF & 0.7019          & \textbf{0.7597} & \textbf{1.3606} & 0.8483          \\ \hline
        DeepMF  & 1.1021          & 0.8036          & 1.5038          & \textbf{0.7797} \\ \hline
        VNCF    & 0.7024          & 0.7857          & 1.4194          & 0.8882          \\ \hline
        NCF     & \textbf{0.6829} & 0.7819          & \textbf{1.3682} & \textbf{0.7766} \\ \hline
        \end{tabular}
        \caption{Mean Squared Error. The lower the better.}
        \label{tab:mse}
    \end{subtable}
     
    \vspace{0.3cm}
     
    \begin{subtable}[h]{\textwidth}
        \centering
        \begin{tabular}{|l|l|l|l|l|}
        \hline
                & FilmTrust       & MovieLens       & MyAnimeList     & Netflix         \\ \hline
        VDeepMF & 0.1794          & \textbf{0.3925} & \textbf{0.4487} & 0.2794          \\ \hline
        DeepMF  & -0.2883         & 0.3574          & 0.3907          & 0.3376          \\ \hline
        VNCF    & 0.1788          & 0.3717          & 0.4248          & 0.2455          \\ \hline
        NCF     & \textbf{0.2017} & 0.3747          & \textbf{0.4456} & \textbf{0.3403} \\ \hline
        \end{tabular}
        \caption{$R^2$ score. The higher the better.}
        \label{tab:r2}
    \end{subtable}
    \caption{Quality of the predictions}
    \label{tab:prediction-error}
\end{table}

\begin{figure}
    \centering
    \begin{subfigure}[b]{0.49\textwidth}
        \centering
        \includegraphics[width=\textwidth]{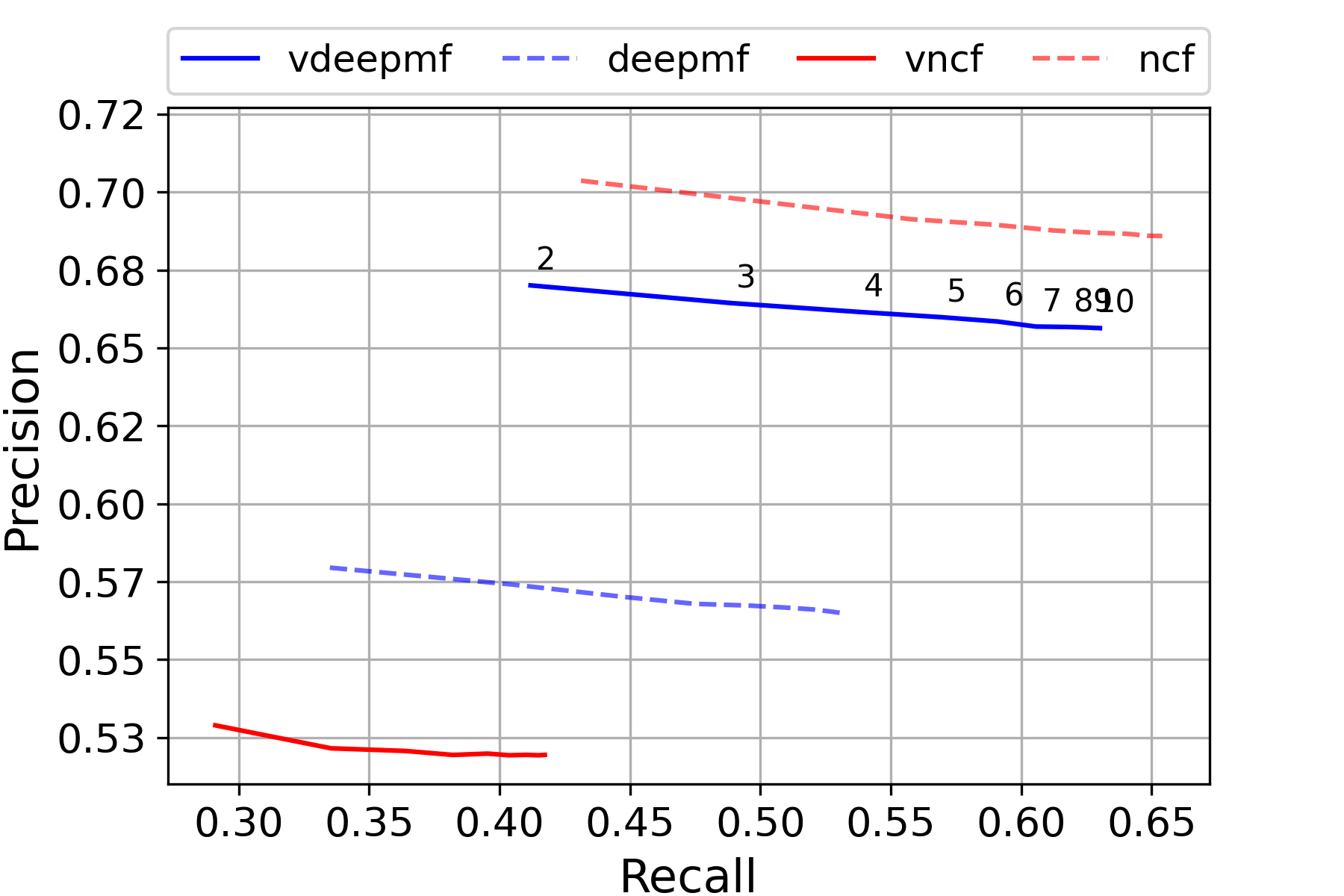}
        \caption{FilmTrust}
        \label{fig:ft-precision-recall}
    \end{subfigure}
    \hfill
    \begin{subfigure}[b]{0.49\textwidth}
        \centering
        \includegraphics[width=\textwidth]{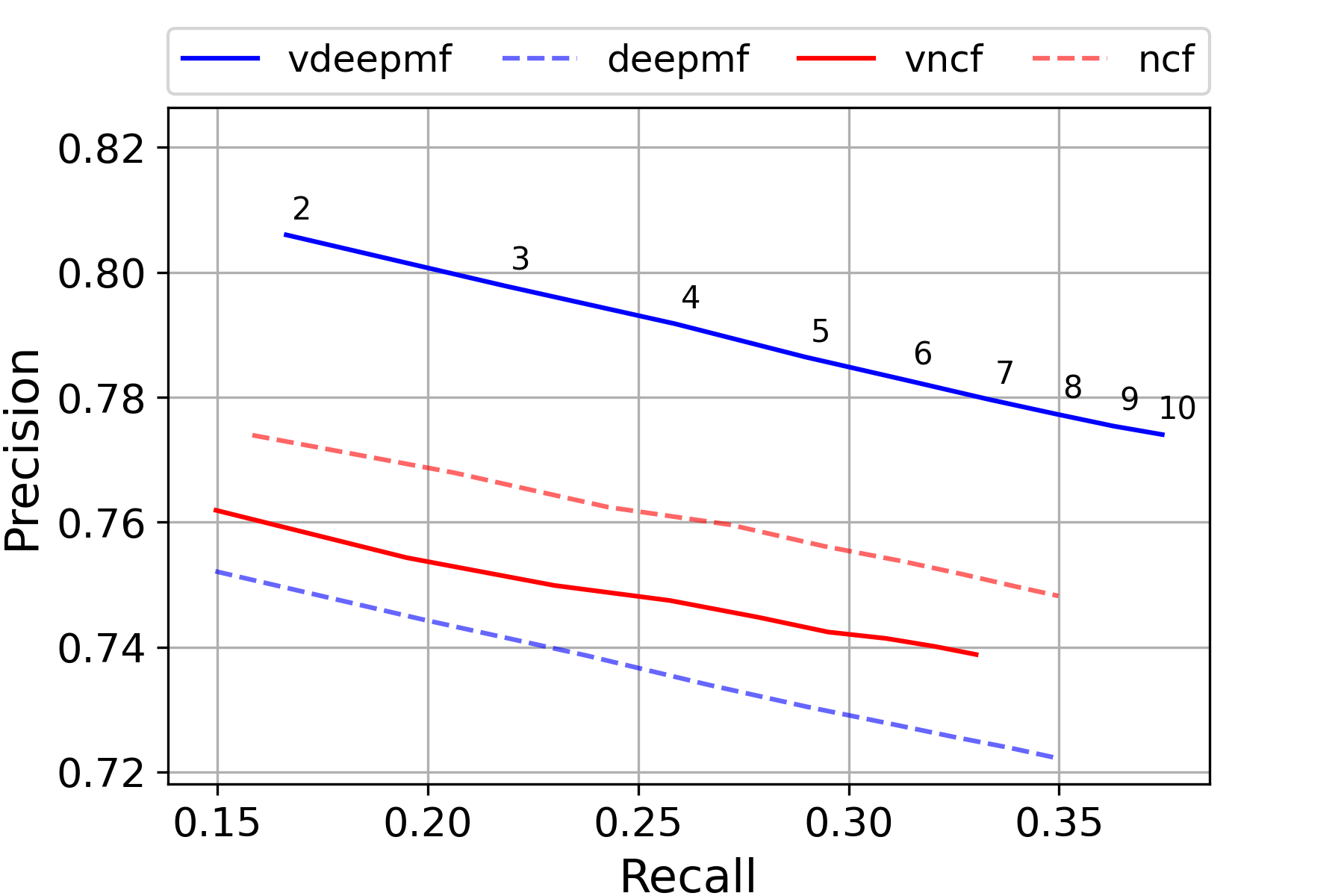}
        \caption{MovieLens}
        \label{fig:ml1m-precision-recall}
    \end{subfigure}
    
    \vspace{0.3cm}
    
    \begin{subfigure}[b]{0.49\textwidth}
        \centering
        \includegraphics[width=\textwidth]{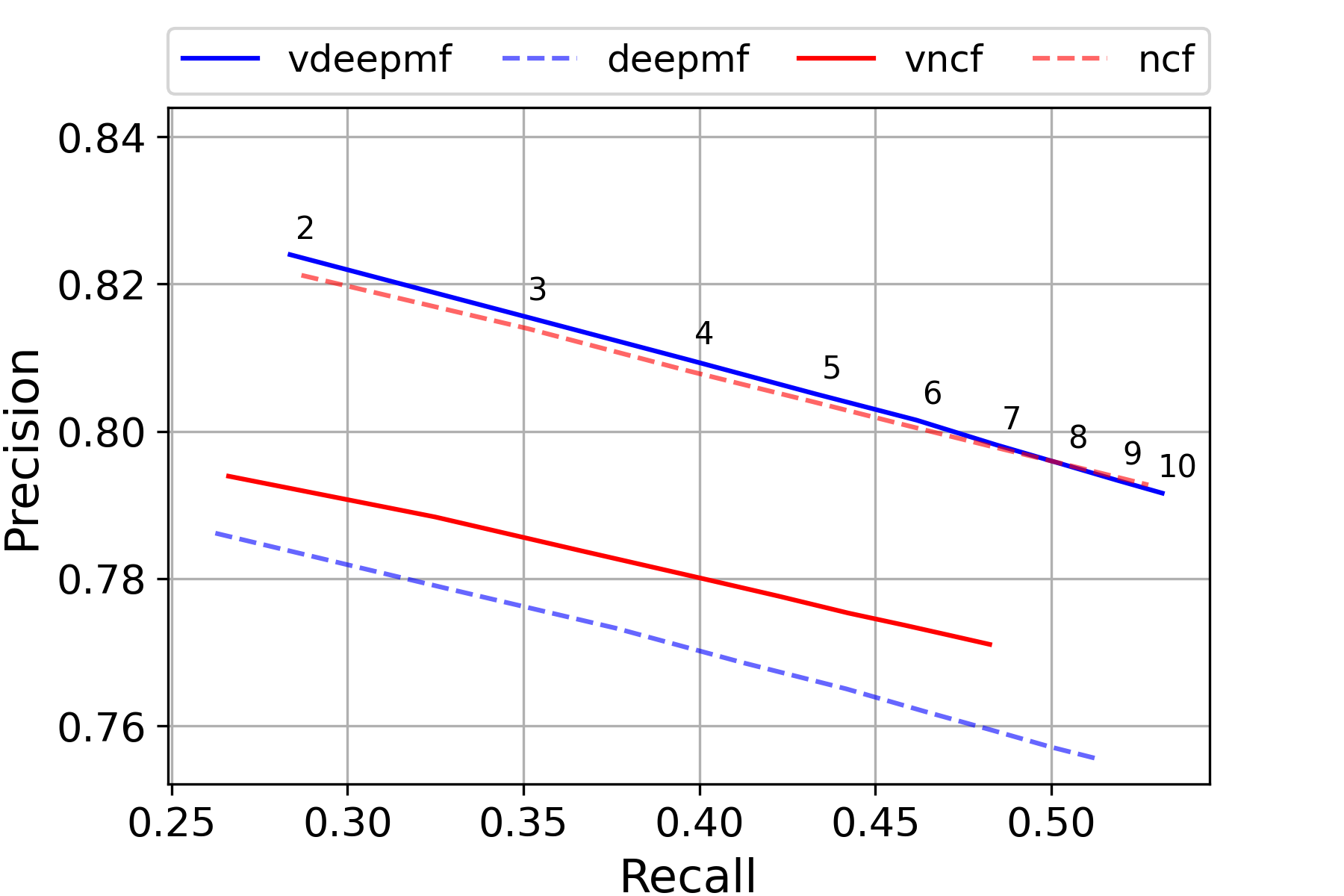}
        \caption{MyAnimeList}
        \label{fig:anime-precision-recall}
    \end{subfigure}
    \hfill
    \begin{subfigure}[b]{0.49\textwidth}
        \centering
        \includegraphics[width=\textwidth]{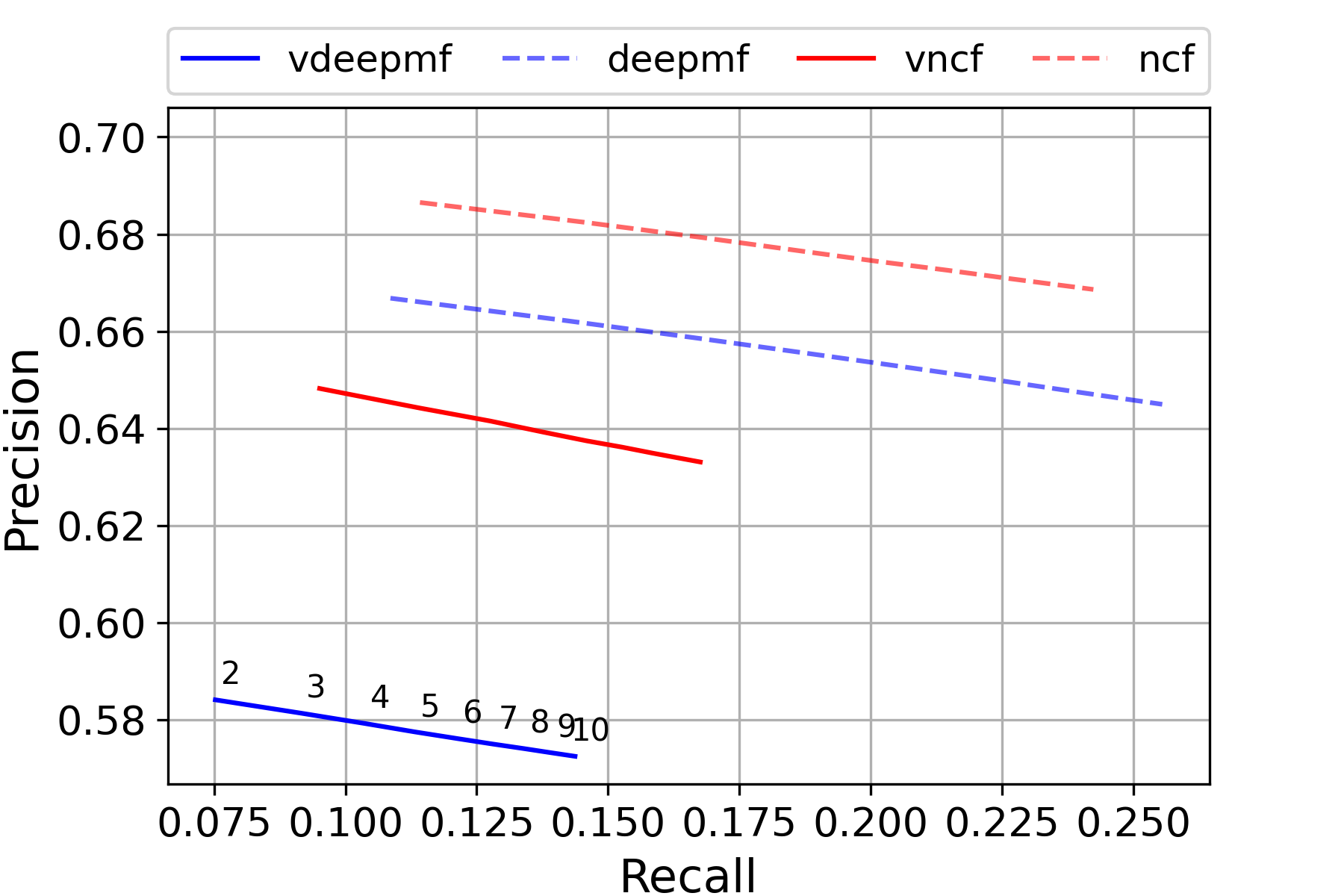}
        \caption{Netflix}
        \label{fig:netflix-precision-recall}
    \end{subfigure}
    
    \caption{Quality of the recommendations measured by \textrm{precision} and \textrm{recall}. The higher the better.}
    \label{fig:precision-recall}
\end{figure}

Furthermore, \Cref{fig:precision-recall} contains the precision and recall results. In FilmTrust (\cref{fig:ft-precision-recall}) we can observe that the proposed variational approach reports a huge benefit for the \ac{DeepMF} model and significantly worsens the results of the \ac{NCF} model. In MovieLens (\cref{fig:ml1m-precision-recall}) and MyAnimeList (\cref{fig:anime-precision-recall}) the same tendency than in FilmTrust is observer, but, in this case, the proposed \ac{VDeepMF} model is the model that computes the best recommendations for these datasets. In Netflix (\cref{fig:netflix-precision-recall}) the proposed variational approach decreases the quality of the recommendations. These results are consistent with those analyzed when measuring the quality of the predictions. Consequently, it is evident that the proposed variational approach works adequately when the dataset is not too large and the model used is not too complex.

\begin{figure}
    \centering
    \begin{subfigure}[b]{0.49\textwidth}
        \centering
        \includegraphics[width=\textwidth]{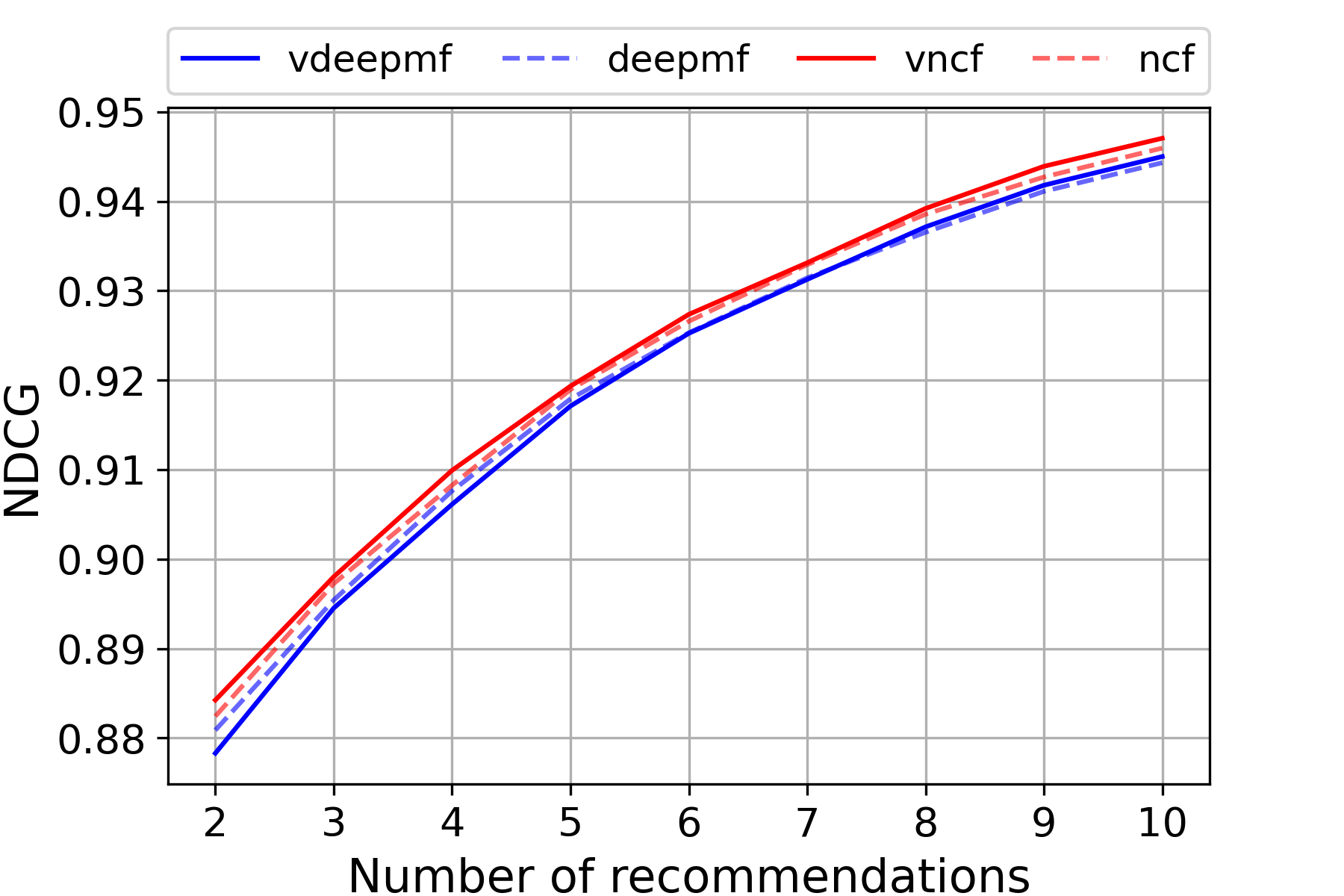}
        \caption{FilmTrust}
        \label{fig:ft-ndcg}
    \end{subfigure}
    \hfill
    \begin{subfigure}[b]{0.49\textwidth}
        \centering
        \includegraphics[width=\textwidth]{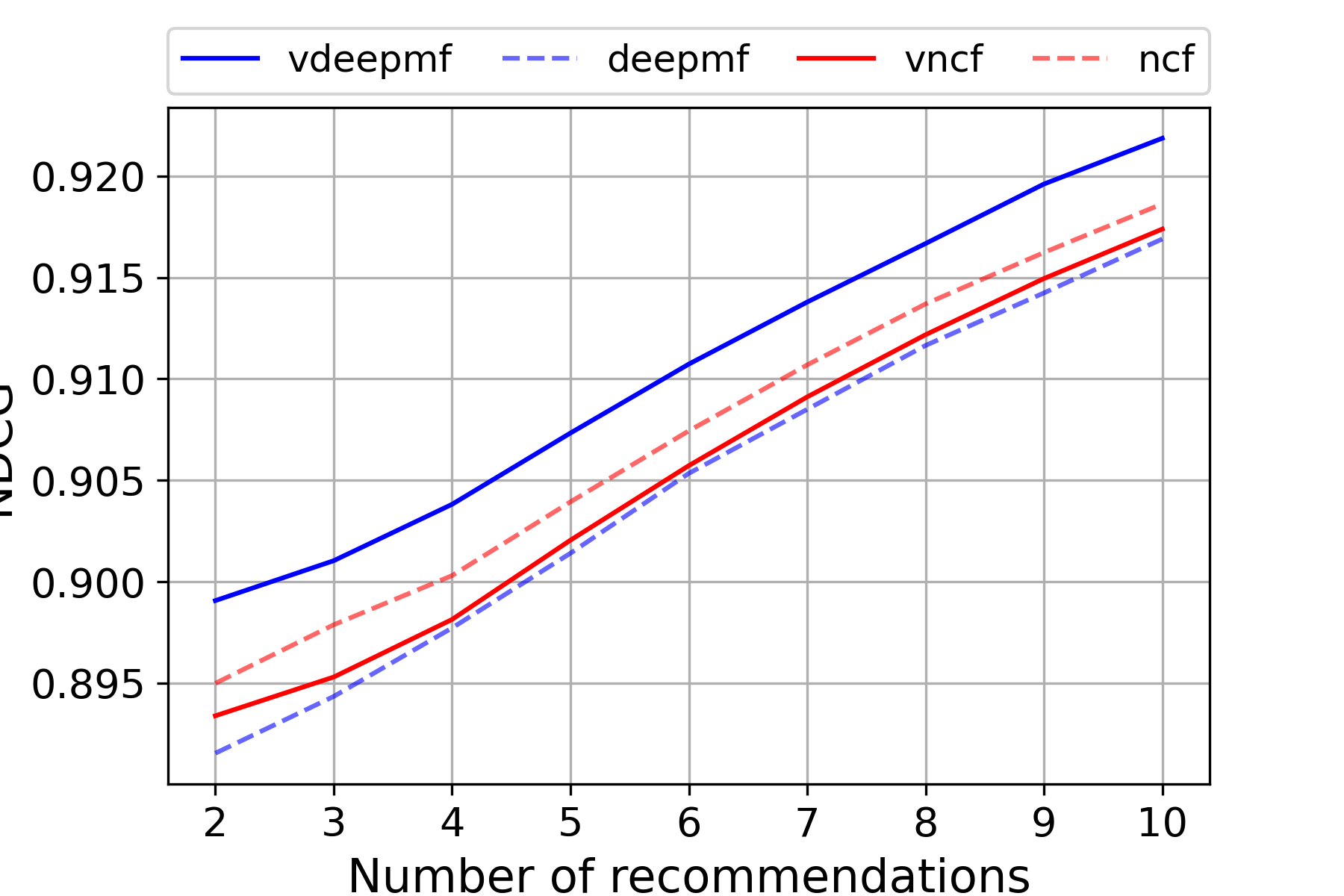}
        \caption{MovieLens}
        \label{fig:ml1m-ndcg}
    \end{subfigure}
    
    \vspace{0.3cm}
    
    \begin{subfigure}[b]{0.49\textwidth}
        \centering
        \includegraphics[width=\textwidth]{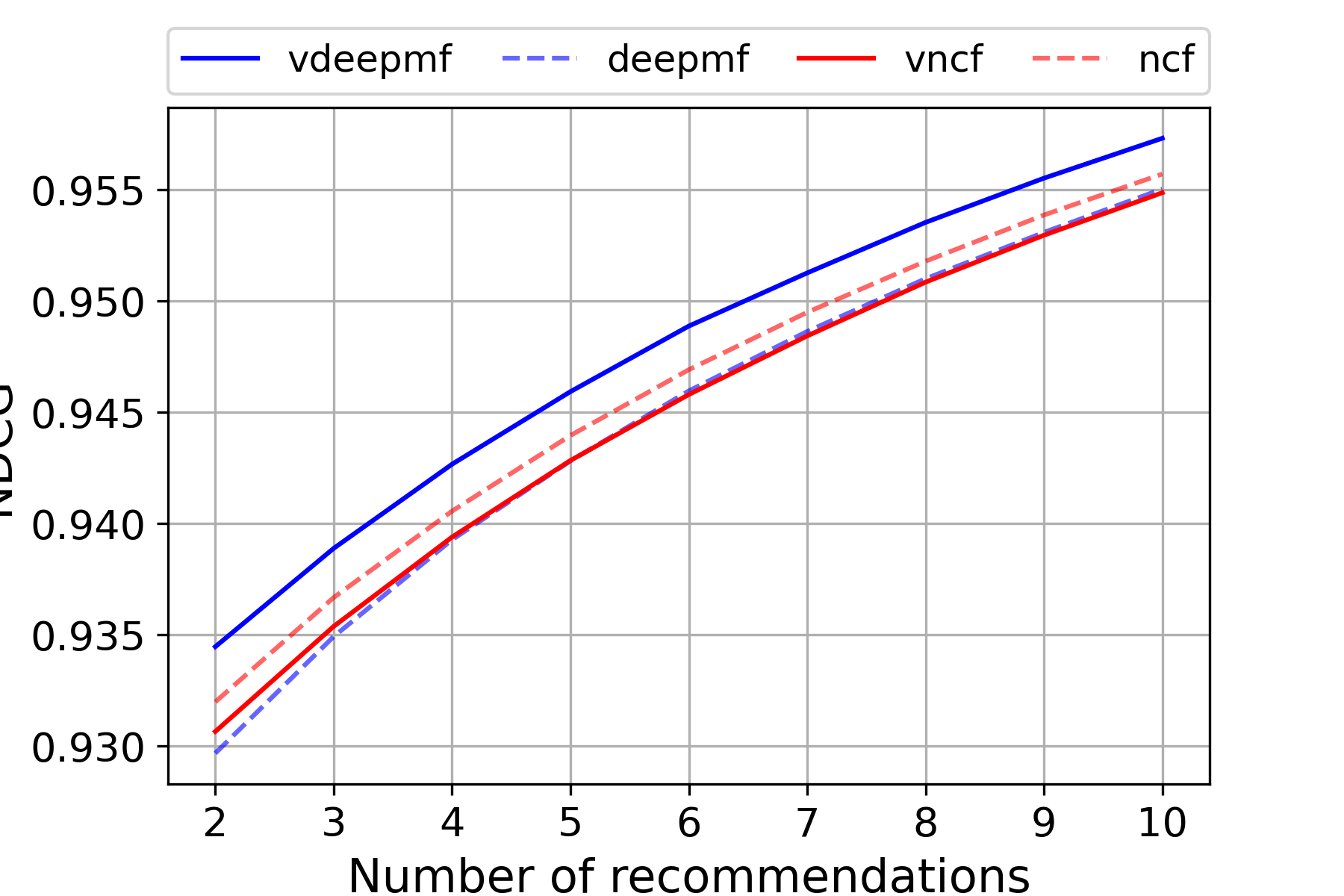}
        \caption{MyAnimeList}
        \label{fig:anime-ndcg}
    \end{subfigure}
    \hfill
    \begin{subfigure}[b]{0.49\textwidth}
        \centering
        \includegraphics[width=\textwidth]{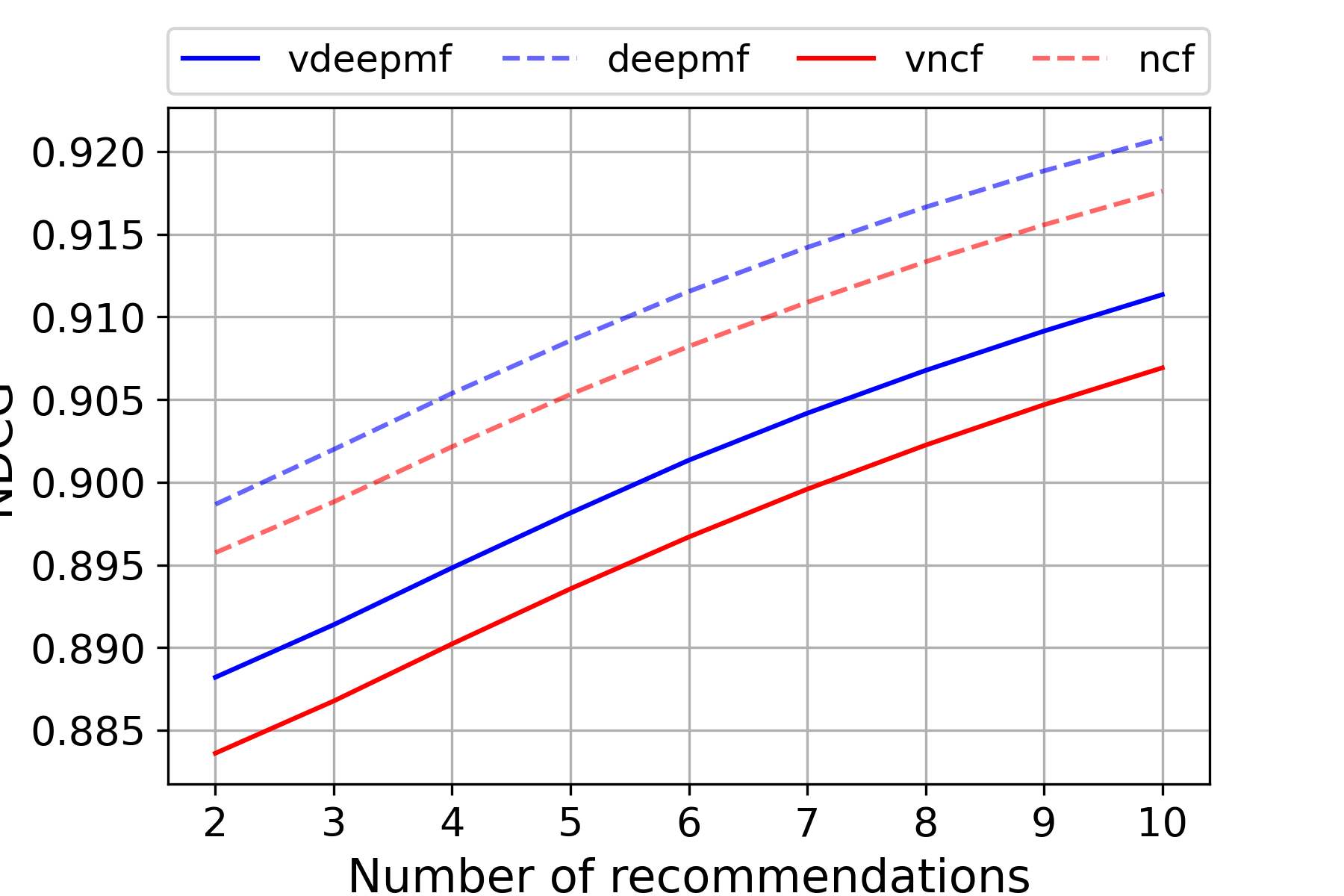}
        \caption{Netflix}
        \label{fig:netflix-ndcg}
    \end{subfigure}
    
    \caption{Quality of the recommendations lists measured by \textrm{NDCG}. The higher the better.}
    \label{fig:ndcg}
\end{figure}

Additionally, \Cref{fig:ndcg} contains the \ac{nDCG} results. From it, we can observe the same trends as those shown in \cref{fig:precision-recall}. In FilmTrust (\cref{fig:ft-ndcg}), the quality of the recommendation lists do not vary independently of whether the variational approach is used or not. In MovieLens (\cref{fig:ml1m-ndcg}) and MyAnimeList (\cref{fig:anime-ndcg}), the combination of the variacional approach with simple modeling such as \ac{DeepMF}, provides the best results. In Netflix (\cref{fig:netflix-ndcg}), the variational approach significantly worsens the quality of the recommendation lists.

Finally, \cref{tab:execution-time} shows the total time and epochs required by each model to be fitted to each dataset using a Quadro RTX 8000 GPU. Best time for each dataset is in bold. We can observe that including a variational layer to the model significantly reduces the required time for fitting. Variational models are able to generate Shannon entropy that is transferred to the regression stage, leading to a more effective training that requires fewer epochs to be fitted. Therefore, the fitting time needed to reach acceptable results is substantially lower. 

\begin{table}[h!]
\begin{tabular}{|l|l|l|l|l|}
\hline
        & FilmTrust               & MovieLens                & MyAnimeList              & Netflix                   \\ \hline
VDeepMF & 61s (15 epochs)         & \textbf{601s (6 epochs)} & \textbf{7629 (9 epochs)} & \textbf{12655 (3 epochs)} \\ \hline
DeepMF  & 75s (25 epochs)         & 677s (10 epochs)         & 13217s (20 epochs)       & 15697 (4 epochs)          \\ \hline
VNCF    & \textbf{35s (7 epochs)} & 1030s (9 epochs)         & 9945 (9 epochs)          & \textbf{12650 (3 epochs)} \\ \hline
NCF     & 56s (15 epochs)         & 876 (10 epochs)          & 12111s (15 epochs)       & 16896 (4 epochs)          \\ \hline
\end{tabular}
\caption{Fitting time using a Quadro RTX 8000.}
\label{tab:execution-time}
\end{table}

\section{Conclusions} 
\label{sec:conclusions}

In the latest trends, accuracy of \acp{RS} is being improved by using deep learning models such as deep matrix factorization and neural collaborative filtering. However, these models do not incorporate stochasticity in their design, unlike variational autoencoders do. Variational random sampling has been used to create augmented input raw data in the collaborative filtering context, but the inherent collaborative filtering data sparsity makes it difficult to get accurate results. This paper applies the variational concept not to generate augmented sparse data, but to create augmented samples in the latent space codified at the dense inner layers of the proposed neural network. This is an innovative approach trying to combine the potential of the variational stochasticity with the augmentation concept. Augmented samples are generated in the dense latent space of the neural network model. In this way, we avoid the sparse scenario in the variational process.

The results show an important improvement when the proposed models are applied to middle-size representative collaborative filtering datasets, compared to the state-of-art baselines, testing both prediction and recommendation quality measures. In contrast, testing on the huge Netflix dataset not only leads to no improvement, but the recommendation quality worsens: increasing Shannon entropy in rich latent spaces causes that the negative effect of the introduced noise exceeds its benefit. Therefore, the proposed deep variational models should be applied for seeking to a fair balance between their positive enrichment and their negative noise injection. The results presented in this work can be considered as generalizable, since they were analyzed in four representative and open \ac{CF} datasets. Researchers can reproduce our experiments and easily create their own models by using the provided framework referenced in section \ref{sec:model}. The authors of this work are committed to reproducible science, so the code used in these experiments is publicly available.

Among the most promising future works, we propose the following: 1) Introducing the variational process in the alternative inner layers of the relevant architectures in the collaborative filtering area, 2) Screening the learning evolution in the training process, since it is faster than the classical models but it also requires early stopping in the training stage, 3) Providing further theoretical explanations of the properties of the \ac{CF} datasets, in terms of Shannon entropy or other statistical features, that ensure a good performance of the proposed models, 4) Applying probabilistic deep learning models in the \ac{CF} field to capture complex non-linear stochastic relationships between random variables, and 5) Testing the impact of the proposed concept when recommendations are made to groups of users.

\section*{Acknowledgements}

\'A. G.-P. acknowledges the hospitality of the Department of Mathematics at Universidad Aut\'onoma de Madrid where part of this work was developed. This work was partially supported by \textit{Ministerio de Ciencia e Innovación} of Spain under the project PID2019-106493RB-I00 (DL-CEMG) and the \textit{Comunidad de Madrid} under \textit{Convenio Plurianual} with the Universidad Politécnica de Madrid in the actuation line of \textit{Programa de Excelencia para el Profesorado Universitario}.

\section*{Conflict of interest}

The authors declare that they have no conflict of interest.

\bibliographystyle{spbasic}      
\bibliography{references}         

\begin{thebibliography}{34}
\providecommand{\natexlab}[1]{#1}
\providecommand{\url}[1]{{#1}}
\providecommand{\urlprefix}{URL }
\expandafter\ifx\csname urlstyle\endcsname\relax
  \providecommand{\doi}[1]{DOI~\discretionary{}{}{}#1}\else
  \providecommand{\doi}{DOI~\discretionary{}{}{}\begingroup
  \urlstyle{rm}\Url}\fi
\providecommand{\eprint}[2][]{\url{#2}}

\bibitem[{Azathoth(2018)}]{myanimelist}
Azathoth (2018) {MyAnimeList Dataset}.
  \url{https://www.kaggle.com/azathoth42/myanimelist}, [Online; accessed
  06-July-2021]

\bibitem[{Beel et~al.(2013)Beel, Langer, Genzmehr, Gipp, Breitinger, and
  N{\"u}rnberger}]{beel2013research}
Beel J, Langer S, Genzmehr M, Gipp B, Breitinger C, N{\"u}rnberger A (2013)
  Research paper recommender system evaluation: a quantitative literature
  survey. In: Proceedings of the International Workshop on Reproducibility and
  Replication in Recommender Systems Evaluation, pp 15--22

\bibitem[{Bennett et~al.(2007)Bennett, Lanning et~al.}]{bennett2007netflix}
Bennett J, Lanning S, et~al. (2007) The netflix prize. In: Proceedings of KDD
  cup and workshop, New York, NY, USA., vol 2007, p~35

\bibitem[{Bobadilla et~al.(2020{\natexlab{a}})Bobadilla, Alonso, and
  Hernando}]{bobadilla2020deep}
Bobadilla J, Alonso S, Hernando A (2020{\natexlab{a}}) Deep learning
  architecture for collaborative filtering recommender systems. Applied
  Sciences 10(7):2441

\bibitem[{Bobadilla et~al.(2020{\natexlab{b}})Bobadilla, Lara-Cabrera,
  Gonz{\'a}lez-Prieto, and Ortega}]{bobadilla2020deepfair}
Bobadilla J, Lara-Cabrera R, Gonz{\'a}lez-Prieto {\'A}, Ortega F
  (2020{\natexlab{b}}) Deepfair: deep learning for improving fairness in
  recommender systems. arXiv preprint arXiv:200605255

\bibitem[{Bobadilla et~al.(2021)Bobadilla, Gonz{\'a}lez-Prieto, Ortega, and
  Lara-Cabrera}]{bobadilla2021deep}
Bobadilla J, Gonz{\'a}lez-Prieto {\'A}, Ortega F, Lara-Cabrera R (2021) Deep
  learning feature selection to unhide demographic recommender systems factors.
  Neural Computing and Applications 33(12):7291--7308

\bibitem[{{\c{C}}ano and Morisio(2017)}]{ccano2017hybrid}
{\c{C}}ano E, Morisio M (2017) Hybrid recommender systems: A systematic
  literature review. Intelligent Data Analysis 21(6):1487--1524

\bibitem[{Chollet et~al.(2015)}]{chollet2015keras}
Chollet F, et~al. (2015) Keras. \url{https://keras.io}

\bibitem[{Deldjoo et~al.(2020)Deldjoo, Schedl, Cremonesi, and
  Pasi}]{deldjoo2020recommender}
Deldjoo Y, Schedl M, Cremonesi P, Pasi G (2020) Recommender systems leveraging
  multimedia content. ACM Computing Surveys (CSUR) 53(5):1--38

\bibitem[{Fan and Cheng(2018)}]{fan2018matrix}
Fan J, Cheng J (2018) Matrix completion by deep matrix factorization. Neural
  Networks 98:34--41

\bibitem[{F{\'e}votte and Idier(2011)}]{fevotte2011algorithms}
F{\'e}votte C, Idier J (2011) Algorithms for nonnegative matrix factorization
  with the $\beta$-divergence. Neural computation 23(9):2421--2456

\bibitem[{Forouzandeh et~al.(2021)Forouzandeh, Berahmand, and
  Rostami}]{forouzandeh2021presentation}
Forouzandeh S, Berahmand K, Rostami M (2021) Presentation of a recommender
  system with ensemble learning and graph embedding: a case on movielens.
  Multimedia Tools and Applications 80(5):7805--7832

\bibitem[{Gao et~al.(2021)Gao, Zhang, Yu, Li, Wen, and
  Xiong}]{gao2021recommender}
Gao M, Zhang J, Yu J, Li J, Wen J, Xiong Q (2021) Recommender systems based on
  generative adversarial networks: A problem-driven perspective. Information
  Sciences 546:1166--1185

\bibitem[{Guo et~al.(2013)Guo, Zhang, and Yorke-Smith}]{guo2013novel}
Guo G, Zhang J, Yorke-Smith N (2013) A novel bayesian similarity measure for
  recommender systems. In: Proceedings of the 23rd International Joint
  Conference on Artificial Intelligence (IJCAI), pp 2619--2625

\bibitem[{Harper and Konstan(2015)}]{harper2015movielens}
Harper FM, Konstan JA (2015) The movielens datasets: History and context. Acm
  transactions on interactive intelligent systems (tiis) 5(4):1--19

\bibitem[{He et~al.(2017)He, Liao, Zhang, Nie, Hu, and Chua}]{he2017neural}
He X, Liao L, Zhang H, Nie L, Hu X, Chua TS (2017) Neural collaborative
  filtering. In: Proceedings of the 26th international conference on world wide
  web, pp 173--182

\bibitem[{Hernando et~al.(2016)Hernando, Bobadilla, and
  Ortega}]{hernando2016non}
Hernando A, Bobadilla J, Ortega F (2016) A non negative matrix factorization
  for collaborative filtering recommender systems based on a bayesian
  probabilistic model. Knowledge-Based Systems 97:188--202

\bibitem[{Kulkarni and Rodd(2020)}]{kulkarni2020context}
Kulkarni S, Rodd SF (2020) Context aware recommendation systems: A review of
  the state of the art techniques. Computer Science Review 37:100255

\bibitem[{Liang et~al.(2018)Liang, Krishnan, Hoffman, and
  Jebara}]{liang2018variational}
Liang D, Krishnan RG, Hoffman MD, Jebara T (2018) Variational autoencoders for
  collaborative filtering. In: Proceedings of the 2018 world wide web
  conference, pp 689--698

\bibitem[{Liu et~al.(2020{\natexlab{a}})Liu, Gherbi, Wei, Li, and
  Cheriet}]{liu2020multispectral}
Liu X, Gherbi A, Wei Z, Li W, Cheriet M (2020{\natexlab{a}}) Multispectral
  image reconstruction from color images using enhanced variational autoencoder
  and generative adversarial network. IEEE Access

\bibitem[{Liu et~al.(2020{\natexlab{b}})Liu, Siu, and Chan}]{liu2020photo}
Liu ZS, Siu WC, Chan YL (2020{\natexlab{b}}) Photo-realistic image
  super-resolution via variational autoencoders. IEEE Transactions on Circuits
  and Systems for Video Technology

\bibitem[{Liu et~al.(2020{\natexlab{c}})Liu, Siu, Wang, Li, and
  Cani}]{liu2020unsupervised}
Liu ZS, Siu WC, Wang LW, Li CT, Cani MP (2020{\natexlab{c}}) Unsupervised real
  image super-resolution via generative variational autoencoder. In:
  Proceedings of the IEEE/CVF Conference on Computer Vision and Pattern
  Recognition Workshops, pp 442--443

\bibitem[{Mnih and Salakhutdinov(2007)}]{mnih2007probabilistic}
Mnih A, Salakhutdinov RR (2007) Probabilistic matrix factorization. Advances in
  neural information processing systems 20:1257--1264

\bibitem[{Narang and Taneja(2018)}]{narang2018deep}
Narang S, Taneja N (2018) Deep content-collaborative recommender system
  (dccrs). In: 2018 International Conference on Advances in Computing,
  Communication Control and Networking (ICACCCN), IEEE, pp 110--116

\bibitem[{Nisha and Mohan(2019)}]{nisha2019social}
Nisha C, Mohan A (2019) A social recommender system using deep architecture and
  network embedding. Applied Intelligence 49(5):1937--1953

\bibitem[{Rendle et~al.(2020)Rendle, Krichene, Zhang, and
  Anderson}]{rendle2020neural}
Rendle S, Krichene W, Zhang L, Anderson J (2020) Neural collaborative filtering
  vs. matrix factorization revisited. In: Fourteenth ACM Conference on
  Recommender Systems, pp 240--248

\bibitem[{Shannon and Weaver(1949)}]{shannon1949mathematical}
Shannon CE, Weaver W (1949) The mathematical theory of communication. Urbana:
  University of Illinois Press 96

\bibitem[{Shokeen and Rana(2020)}]{shokeen2020study}
Shokeen J, Rana C (2020) A study on features of social recommender systems.
  Artificial Intelligence Review 53(2):965--988

\bibitem[{Trigeorgis et~al.(2016)Trigeorgis, Bousmalis, Zafeiriou, and
  Schuller}]{trigeorgis2016deep}
Trigeorgis G, Bousmalis K, Zafeiriou S, Schuller BW (2016) A deep matrix
  factorization method for learning attribute representations. IEEE
  transactions on pattern analysis and machine intelligence 39(3):417--429

\bibitem[{Wan et~al.(2020)Wan, Xia, Kong, Hsu, Huang, and Ma}]{wan2020deep}
Wan L, Xia F, Kong X, Hsu CH, Huang R, Ma J (2020) Deep matrix factorization
  for trust-aware recommendation in social networks. IEEE Transactions on
  Network Science and Engineering 8(1):511--528

\bibitem[{Wen et~al.(2018)Wen, She, Li, and Mao}]{wen2018visual}
Wen J, She J, Li X, Mao H (2018) Visual background recommendation for dance
  performances using deep matrix factorization. ACM Transactions on Multimedia
  Computing, Communications, and Applications (TOMM) 14(1):1--19

\bibitem[{Xue et~al.(2017)Xue, Dai, Zhang, Huang, and Chen}]{xue2017deep}
Xue HJ, Dai X, Zhang J, Huang S, Chen J (2017) Deep matrix factorization models
  for recommender systems. In: IJCAI, Melbourne, Australia, vol~17, pp
  3203--3209

\bibitem[{Zhang et~al.(2021)Zhang, Liu, Zuo, Lu, and Lian}]{zhang2021online}
Zhang Ss, Liu Jw, Zuo X, Lu Rk, Lian Sm (2021) Online deep learning based on
  auto-encoder. Applied Intelligence pp 1--20

\bibitem[{Zou et~al.(2020)Zou, Chen, He, Li, Zhang, and Gan}]{zou2020ndmf}
Zou G, Chen J, He Q, Li KC, Zhang B, Gan Y (2020) Ndmf: Neighborhood-integrated
  deep matrix factorization for service qos prediction. IEEE Transactions on
  Network and Service Management 17(4):2717--2730

\end{thebibliography}

\end{document}